\documentclass[aps,prd,preprintnumbers,nofootinbib,onecolumn]{revtex4}
\usepackage{amsmath,amsfonts,amssymb}
\usepackage{graphicx,epsfig}
\usepackage{float}
\usepackage{color}
\usepackage{amsthm}
\usepackage[english]{babel}
\usepackage{hyperref}

\begin{document}
\title{Branes in Gravity's Rainbow}

\author{Amani Ashour}\email{amani87.math@gmail.com}\affiliation{Mathematics Department, Faculty of science, Damascus university, Damascus, Syria}
\author{Mir Faizal}\email{f2mir@uwaterloo.ca}\affiliation{Department of Physics and Astronomy, University of Lethbridge,
Lethbridge, Alberta, T1K 3M4, Canada}
\author{Ahmed Farag Ali}\email{ahmed.ali@fsc.bu.edu.eg}\affiliation{Department of Physics, Faculty of Science, Benha University, Benha, 13518, Egypt}
\author{Fay\c{c}al Hammad}\email{fhammad@ubishops.ca}\affiliation{Physics Department STAR Research Cluster, Bishop's University, and Physics Department, Champlain College-Lennoxville, 2600 College Street, Sherbrooke, Qu\'{e}bec J1M 1Z7, Canada}

\begin{abstract}
In this work, we investigate the thermodynamics of black $p$-branes (BB) in the context of Gravity's Rainbow.
We investigate this using  rainbow functions that have been motivated from loop quantum gravity   and   $\kappa$-Minkowski
noncommutative spacetime.  
Then  for the sake of comparison, we examine
a couple of other rainbow functions that have also appeared in the literature. We show that, for consistency, Gravity's
Rainbow  imposes a constraint on the minimum mass of the BB, a constraint that we interpret here as implying the
existence of a black $p$-brane remnant. This interpretation is supported by the computation of the black $p$-brane's heat
capacity that shows that the latter vanishes when the Schwarzschild radius takes on a value that is bigger than its extremal limit.
We found that the same conclusion is reached for the third version of rainbow functions treated here but not with the second one
for which only standard black $p$-brane thermodynamics is recovered.
\end{abstract}

\maketitle

\section{Introduction}\label{Sec:1}
One common feature among most of semi-classical approaches to quantum gravity
is a Lorentz invariance violation due to a departure from the usual relativistic dispersion relation caused by a redefinition of the
physical momentum and physical energy at the Planck scale. The source of this departure comes from many approaches
such as loop quantum gravity  \cite{AmelinoCamelia:1996pj, amerev} spacetime discreteness~\cite{'tHooft:1996uc},
spontaneous symmetry breaking of Lorentz invariance in string field theory~\cite{LIstring},
spacetime foam models~\cite{AmelinoCamelia:1997gz} and spin-networks~\cite{Gambini:1998it}.
A more recent approach that also predicts Lorentz invariance violation is non-commutative geometry~\cite{Carroll:2001ws}.
All these findings suggest that Lorentz violation might be a generic and an essential property when it comes to constructing
a quantum theory of gravity. Mathematically, the departure from Lorentz invariance is expressed in the form of a modified dispersion
relation (MDR). This modification could be behind anomalies that might occur in ultra-high energy cosmic rays and
TeV photons~\cite{AmelinoCamelia:1997gz,AmelinoCamelia:1997jx,AmelinoCamelia:1999wk}. Modern observations are
recently gaining the needed sensitivity to measure such effects, and are expected to be further improved in the coming
few years\footnote{Threshold anomalies are only predicted by MDR scenarios with a preferred reference frame in which they imply
a full violation of relativistic symmetries. These anomalies are, however, not predicted by scenarios in which MDR is due to
a deformation of relativistic symmetry with no preferred reference frame~\cite{AmelinoCamelia:2002dx}.}. For a recent detailed
review of MDR theories and the possibility of getting physical observations, we refer the reader to Ref.~\cite{amerev}.

The theory that naturally produces the MDR is the so-called doubly special relativity (DSR) \cite{AmelinoCamelia:2000mn}.
DSR is considered as an extension of the special theory of relativity that extends the invariant quantities to include the Planck energy scale besides the speed of light. The simplest realization of the DSR is based on a non-linear Lorentz transformation in momentum space. This non-linear transformation implies a deformed Lorentz symmetry such that the usual dispersion relations of special relativity become modified by corrections relevant only at the Planck scale. It should be mentioned that Lorentz invariance violation and Lorentz invariance deformation are in general conceptually different scenarios. Here we shall adopt Lorentz invariance deformation by considering DSR and its extension in models of Gravity's Rainbow.

In the framework of DSR, the definition of the dual position space suffers a nonlinearity of the Lorentz transformation. To resolve this issue,
Magueijo and Smolin \cite{Magueijo:2002xx} proposed a doubly general relativity in which one assumes that the spacetime background felt by a test particle
depends on the energy of the latter. Therefore, there will not be a single metric describing the spacetime as seen by test particles, but a one-parameter
family of metrics that depends on the energy and momentum of these test particles, forming in a sense a 'rainbow' of metrics or geometries. This idea is
usually known as Gravity's Rainbow (and sometimes Rainbow Gravity). This is based on new Lorentz transformations \cite{AmelinoCamelia:1996pj},
which lead to modified dispersion relation. It may be noted that a
modified equivalence principle has been   proposed in Ref.~\cite{Magueijo:2002xx}, and this has lead to the development of gravity's rainbow.
Gravity's rainbow depends on the rainbow function chosen,  and potential investigations based on Gravity's Rainbow can be found in Ref.~\cite{Galan:2004st}.
The choice of the rainbow functions (denoted by $f(E/E_{p})$ and $g(E/E_{p})$) is important for making physical predictions.
Based on different arbitrary choices of these functions, many aspects of Gravity's rainbow have been applied in
Ref.~\cite{Galan:2004st,FRWRainbow} to black hole physics using the Schwarzschild metric, and to inflation and
its predicted scale-invariant fluctuations using the Friedmann-Lema\^{\i}tre-Robertson-Walker metric in Ref.~\cite{Barrow:2013gia}. Furthermore,
as more recent application of Rainbow Gravity, a recent investigation on the possibility of resolving the Big Bang singularity was
carried out in Ref.~\cite{Awad:2013nxa}.

In this letter, we continue the investigation of the effects of Gravity's Rainbow on Black hole thermodynamics.
We use the modified dispersion relation of Eq.~(\ref{MDR}), which fixes the rainbow functions $f(E/E_{p})$ and $g(E/E_{p})$
and use them to study the black $p$-brane thermodynamics and investigate its new properties. The motivation for investigating $p$-brane thermodynamics within the framework of Gravity's rainbow comes from the fact that recently intense investigation in string theory has been conducted in the UV sector of gravity which, in turn, turns out to be related to Gravity's Rainbow.


In fact, in order to obtain a UV completion of gravity, such that it reduces to General Relativity in the IR limit, space and time are made to have
different Lifshitz scaling \cite{HoravaPRD,HoravaPRL}. This approach for obtaining the UV completion of gravity is called the
Horava-Lifshitz gravity. It is such a UV completion that has been recently much studied in the context of string theory. Indeed,
a UV completion (by taking a different Lifshitz scaling for space and time ) has been studied in the context of
type IIA string theory \cite{A}, type  IIB string theory \cite{B}, AdS/CFT correspondence
\cite{ho, h1, h2,  oh}, dilaton black branes \cite{d, d1}, and dilaton black holes \cite{dh, hd}. It turns out,
however, that there is another way to obtain a UV completion of General Relativity, and this approach is non other than
the Gravity's Rainbow \cite{Magueijo:2002xx} we have discussed above.

Gravity's Rainbow is actually related to Horava-Lifshitz gravity \cite{re}. This is because both these UV completions of General Relativity are based on the  modification of the usual energy-momentum dispersion relation in the UV limit, such that it reduces to the usual energy-momentum dispersion relation in the IR limit. Furthermore, such a modification of the energy-momentum relation also occurs in ghost condensation \cite{FaizalJPA} and non-commutative geometry \cite{Carroll,FaizalMPLA}. It may be noted that non-commutative geometry occurs due to background fluxes in string theory \cite{st, st1}. Noncommutative geometry is also
used to derive one of the most important rainbow functions in Gravity's Rainbow
\cite{Amelino,Jacob}. In this paper,  we will use this rainbow function which is motivated by noncommutative geometry. As noncommutative geometry
occurs due to background fluxes in string theory, this rainbow deformation can be thought to be dual to some background fluxes in the string theory. Furthermore, as the UV completion of such  geometries has already been studied using the formalism of Horava-Lifshitz gravity,
and as Horava-Lifshitz gravity is related to Gravity's Rainbow \cite{re}, it is important to study the rainbow deformation of such geometries.

In fact, there are other motivations to study the deformation of the usual energy-momentum relation in string theory. This is because it is possible for a tachyon field to have the wrong sign for its mass squared in string field theory. The existence of such a tachyon field can make the
perturbative string vacuum become unstable \cite{58}. This, in turn, can spontaneously break the Lorentz symmetry,
and such a spontaneous breaking of the Lorentz symmetry will deform  the usual energy-momentum
relations.  The spontaneous breaking of the Lorentz symmetry occurs due to the
gravitational Higgs mechanism in supergravity theories, and this, again, deforms the usual energy-momentum relation \cite{59}. Hence, there is a good physical motivation to study the deformation of the energy-momentum relation in the UV limit within string theory.
Such deformation of geometries which occur in string theory have usually been studied using the formalism of Horava-Lifshitz gravity.
However, as Horava-Lifshitz gravity is related to Gravity's Rainbow, it is important to study such UV completion of such geometries using
Gravity's Rainbow. So, in this paper, we will use a rainbow function, which has been motivated from noncommutative
geometry to analyze the consequences of UV completion of $p$-branes.

It may be noted that the modification of the usual energy-momentum relation
in the UV limit has also been motivated from the study of cosmic rays \cite{q5, q6}.
Furthermore, experimental tests to confirm the existence of such a modified dispersion relation
have also been suggested \cite{q1}. So, future experiments will either verify or falsify the existence of such dispersion relations  and approaches like
the Horava-Lifshitz gravity and gravity's rainbow.
It may be noted that even though geometry is energy- dependent, such effect only occurs near the Planck scale, where quantum gravitational effects are expected
to modify the
semi-classical geometry in very exotic ways. At low energies such effects can be neglected, and hence, in the IR limit the usual general relativity is recovered. It is worth mentioning that the energy-dependent metric
has been obtained in different approaches to quantum gravity such as low-energy effective field theories in \cite{Lafrance:1994in} and in studying string theory at short distances beyond the Planck scale \cite{Mende:1992pm,Gross:1987ar}.


Now, it turns out that, due to these rainbow functions, an end-point of Hawking radiation is not catastrophic anymore.
Indeed, we found that there should be a black $p$-brane remnant because the specific heat vanishes at some Schwarzschild radius $r_{0}$
greater than its extremal limit and, hence, the black hole could no longer exchange heat with the surrounding space.
The same conclusion applies with a second category of rainbow functions but not with a third. A detailed discussion on this subtle point will be attempted in the conclusion section.

The paper is organized as follows. In Sec.~\ref{Sec:2}, we briefly recall the main steps in obtaining the black $p$-brane's thermodynamics.
In Sec.~\ref{sec:3}, we investigate how Rainbow Gravity effects the BB thermodynamics with the most studied rainbow functions.
A couple of other cases of rainbow functions will be examined in Sec.~\ref{sec:4}. We end this letter with a brief conclusion section.

\section{$p$-Brane Thermodynamics Review}\label{Sec:2}
Let us first briefly review in this section the standard steps followed in finding the thermodynamics of
black $p$-branes in the near-horizon limit \cite{Gubser:1998ex, Polchinski:1996na, Horowitz:1996nw, Harmark}.
The metric in a near-extremal black $p$-brane is given by
\begin{equation}\label{BBmetric}
ds^2=\chi^{-1/2}\left[-\left(1-\frac{r_{0}^{n}}{r^{n}}\right)\mathrm{d}t^{2}+\mathrm{d}y^{i}\mathrm{d}y_{i}\right]
+\chi^{1/2}\left[\left(1-\frac{r_{0}^{n}}{r^{n}}\right)^{-1}\mathrm{d}r^{2}+r^{2}\mathrm{d}\Omega_{n+1}\right],
\end{equation}
where $r_{0}$ is the radial location of the horizon which, when it vanishes, corresponds to the extremal $p$-brane case. $\Omega_{n+1}=2\pi^{1+n/2}/\Gamma(1+n/2)$ is the volume of the unit $(n+1)$-sphere, $y^{i}$ denote the $p=7-n$ spatial coordinates along the brane, which are assumed compactified on a large torus of volume $V$, and
\begin{equation}
\chi=1+\frac{r_{0}^{n}}{r^{n}}\sinh^{2}\alpha.
\end{equation}
$\alpha$ is a dimensionless parameter related to the charge of the brane, $r_{0}$ is the Schwarzschild radius, while the dilaton $\phi$ is related to $\chi$ by $e^{2\phi}=\chi^{(n-4)/2}$. The dilaton gives the relation between the string metric (\ref{BBmetric}) and the Einstein metric $\tilde{g}_{\mu\nu}$ as $\mathrm{d}s^2=e^{2\phi}\tilde{g}_{\mu\nu}\mathrm{d}x^{\mu}\mathrm{d}x^{\nu}$.

Now, as usual, the Hawking temperature \cite{Hawking:1974sw} of a black $p$-brane is also defined by the corresponding surface gravity $\kappa$ as $T=\kappa/2\pi$. Since for any static and spherically symmetric metric $g_{\mu\nu}\mathrm{d}x^{\mu}\mathrm{d}x^{\nu}$ the surface gravity $\kappa$ is given by \cite{GRbooks} $\kappa^{2}=-\frac{1}{4}g^{rr}g^{tt}(\partial_{r}g_{tt})^{2}\big|_{r=r_{0}}$, one easily finds, using the metric (\ref{BBmetric}), that the black $p$-brane temperature reads
\begin{eqnarray}
T=\frac{n}{4\pi r_{0}\cosh\alpha}.\label{hawkT}
\end{eqnarray}
The Bekenstein-Hawking entropy is then obtained by integrating the first law of thermodynamics $\mathrm{d}M=T\mathrm{d}S+\mu\mathrm{d}Q$ \cite{Harmark},
where $\mu=\tanh\alpha$ is the chemical potential, after keeping the charge constant when varying the ADM mass $M$ of the black $p$-brane, related
to the radius $r_{0}$ by (see e.g. \cite{Lu,Caio,Harmark})
\begin{eqnarray}
M=\frac{V\Omega_{n+1}r_{0}^{n}}{16\pi G}\left(n+1+n\sinh\alpha\right)\label{BraneMass}.
\end{eqnarray}
A straightforward calculation using (\ref{hawkT}) and (\ref{BraneMass}) gives,
up to an integration constant, the entropy $S=(4G)^{-1}V\Omega_{n+1}r_0^{n+1}\cosh\alpha$ for the black $p$-brane. Then, using the heat capacity formula $C=T\partial_{r}S/\partial_{r}T$, one easily finds the following heat capacity of standard $p$-branes: $C=-(n+1)(4G)^{-1}V\Omega_{n+1}r_0^{n+1}$. After this brief review of black $p$-branes thermodynamics, we shall examine in the next section how the latter is modified when Gravity's Rainbow is taken into consideration.
\section{Rainbow $p$-Brane Thermodynamics}\label{sec:3}
One of the most interesting classes of MDRs has been suggested in Ref.~\cite{AmelinoCamelia:1996pj, amerev}.
For a particle of mass $m$, energy $E$ and momentum $\vec{p}$, the relation takes at high-energy regimes the following form:
\begin{eqnarray}
m^2=E^{2}-{\vec{p}}^{\,2}+\gamma\,{\vec{p}}^{\,2}\left(\frac{E}{E_{p}}\right)^{q} \label{MDR},
\end{eqnarray}
where $E_{p}$ represents the energy scale at which the usual relativistic dispersion relation is modified. This scale
is naturally taken to be the Planck energy~\cite{AmelinoCamelia:1996pj}. $q$ and $\gamma$ represent, respectively, a
positive integer characterizing the degree of departure from Lorentz invariance, and a free parameter indicating how
strong the deformation manifests itself for energies close to the Planck scale $E_{p}$. This formula is compatible with some of the
results obtained within the loop quantum gravity approach and reflects some results obtained within the framework of $\kappa$-Minkowski
noncommutative spacetime \cite{amerev}. For a discussion on the phenomenological implications of the relation (\ref{MDR}), we encourage
the reader to refer to the discussion   in the detailed review \cite{amerev}.

The starting point in Gravity's Rainbow is the non-linearity of the new Lorentz transformations \cite{AmelinoCamelia:1996pj},
which lead to the following more general modified dispersion relation
\begin{equation}
E^{2}f(E/E_{p})^{2}- \vec{p}^{\,2}g(E/E_{p})^2= m^2, \label{RainbowDisper}
\end{equation}
where $E_{p}$ is, as in relation (\ref{MDR}), the Planck energy scale, $m$ is the mass of the test particle,
and $f(E/E_{p})$ and $g(E/E_{p})$ are commonly known as rainbow functions. In order to recover the usual relativistic
dispersion relation in the low-energy limit, these functions should satisfy $\lim_{E\to 0}f(E/E_{p})=1$ and $\lim_{E\to 0}g(E/E_{p})=1$.
A modified equivalence principle was consequently proposed in Ref.~\cite{Magueijo:2002xx}, according to which a one-parameter family of
energy-dependent orthonormal frame fields give rise to a one-parameter family of energy-dependent metrics:
\begin{equation}
g^{\mu\nu}(E/E_{p})=e_{a}^{\mu}(E/E_{p})e^{a\nu}(E/E_{p}),
\end{equation}
where the new tetrad fields are related to the usual low-energy frame fields $\tilde{e}_{a}^{\mu}$ of general
relativity by the rainbow functions as $f(E/E_{p})e_{0}^{\mu}(E/E_{p})=\tilde{e}_0^{\mu}$ and $g(E/E_{p})e_{i}^{\mu}(E/E_{p})=\tilde{e}_{i}^{\mu}$,
where $i$ is the spatial index, such that in the limit $E/E_{p}\rightarrow0$ one recovers general relativity. By defining also a one-parameter family
of energy-momentum tensors, Einstein equations get modified to
\begin{equation}
G_{\mu\nu}(E/E_{p})=8\pi G(E/E_{p})T_{\mu\nu}(E/E_{p})+g_{\mu\nu}\Lambda(E/E_{p}). \label{RFE}
\end{equation}
Let us now use the modified dispersion relation (\ref{MDR}), motivated from
loop quantum gravity   and   $\kappa$-Minkowski
noncommutative spacetime~\cite{AmelinoCamelia:1997gz,AmelinoCamelia:1996pj},
to deduce the rainbow functions in Eq. (\ref{RainbowDisper}). A simple comparison reveals that
\begin{eqnarray}
f(E/E_{p})=1~~~~\mathrm{and}~~~~g(E/E_{p})=\sqrt{1-\gamma\left(\frac{E}{E_{p}}\right)^q}. \label{RainFunc}
\end{eqnarray}

These rainbow functions will induce a rainbow geometry. Therefore, the non-rotating and non-extremal black $p$-brane metric (\ref{BBmetric}) should accordingly be modified to acquire the following expression \cite{Magueijo:2002xx}:
\begin{eqnarray}\label{rainbowmetric}
ds^2=\frac{\chi^{-1/2}}{f^{2}}\left[-\left(1-\frac{r_{0}^{n}}{r^{n}}\right)\mathrm{d}t^{2}
+\left(\frac{f}{g}\right)^{2}\mathrm{d}y^{i}\mathrm{d}y_{i}\right]
+\frac{\chi^{1/2}}{g^{2}}\left[\left(1-\frac{r_{0}^{n}}{r^{n}}\right)^{-1}\mathrm{d}r^{2}+r^{2}\mathrm{d}\Omega_{n+1}\right].
\end{eqnarray}
Now with this rainbow metric, the Hawking temperature (\ref{hawkT}) will also be modified. Indeed, after computing the surface gravity $\kappa$ corresponding to this metric, using the above recalled definition, one finds a new temperature $T'$ related to the previous temperature by $T'=f^{-1}(E/E_{p})g(E/E_{p})T$. More explicitly, we have
\begin{equation}
T^{\prime}=\frac{n}{4\pi r_{0}\cosh\alpha}\sqrt{1-\gamma\left(\frac{E}{E_{p}}\right)^q}. \label{Rainsur}
\end{equation}
Note that for $f(E/E_{p})=g(E/E_{p})=1$, the new temperature reduces to standard black $p$-brane temperature.

Before proceeding to the calculation of the entropy, let us pause here for a moment and consider this last result in the light of the uncertainty principle as applied to particles near a black hole's horizon of radius $r_{0}$. As it is familiar in black hole physics \cite{Eliasentropy0,Eliasentropy1,Adler,Cavaglia:2003qk}, one appeals at this point to Heisenberg's uncertainty principle, $\Delta x\Delta p\geq1$, that constrains the uncertainties on the position $x$ and momentum $p$ of a quantum particle. One then trades momentum for energy \cite{AmelinoCamelia:2004xx0}, to obtain the lower-bound $E\geq1/\Delta x$ on the energy of the particle. Furthermore, for a particle moving in the near-horizon geometry, one also usually chooses the uncertainty on its position $\Delta x$ to be of the order of the Schwarzschild radius $r_{0}$ \cite{Eliasentropy0,Eliasentropy1,Adler,Cavaglia:2003qk}, and ends up with a lower bound for the energy of the particle given by $E\geq r_{0}^{-1}$. By substituting this lower bound of energy into the modified Hawking Temperature in Eq.~(\ref{Rainsur}), we get:
\begin{equation}
T^{\prime}=\frac{n}{4\pi r_{0}\cosh\alpha}\sqrt{1-\frac{\gamma}{(E_{p}r_{0})^{q}}}.\label{modT}
\end{equation}
It is clear from Eq.~(\ref{modT}) that when setting $\gamma=0$ or assuming that $E/E_{p}\to 0$ (or equivalently, that $E_{p}\to\infty$), one recovers back the standard Hawking temperature of Eq.~(\ref{hawkT}). It is also clear from Eq.~(\ref{modT}) that the modified Hawking temperature would be physical, i.e. real, as long as the location $r_{0}$ of the black $p$-brane's horizon satisfies the inequality $r_0\geq\gamma^{1/q}E_p^{-1}$. In terms of the mass of the black $p$-brane this translates into the following constraint:
\begin{equation}\label{MinMass}
M\geq M_{min}\sim\gamma^{1/q}E_p.
\end{equation}
This constraint can be interpreted as an existence of a black $p$-brane remnant due to Rainbow Gravity.

Let us now find the rainbow black $p$-brane entropy $S'$ and its heat capacity $C$ by using, respectively,
the formulas $T'\mathrm{d}S'=\mathrm{d}M$ and $C'=T'\partial S'/\partial T'$. With the mass $M$
as given in Eq.~(\ref{BraneMass}) and the new temperature $T'$ as given by Eq.~(\ref{modT}), we find the following integral form for entropy:
\begin{equation}\label{modifiedS}
S^{\prime}=\int\frac{\mathrm{d}M}{T'}=\frac{(n+1)V\Omega_{n+1}\cosh\alpha}{4G}\int\frac{r_0^{n}\mathrm{d}r_0}{\sqrt{1-\gamma( E_{p}r_{0})^{-q}}}.
\end{equation}
Let us compute this integral by choosing a concrete example of a $p$-brane such as a $5$-brane, i.e. let us choose $n=2$.
Then let us examine the consequences of the modified dispersion relation (\ref{MDR}) with two different values of the integer $q$.
Let us first start with a value that is not greater than $n$. The simplest result for the above integral is obtained for $q=n=2$ because
the form of the entropy for the case $q=1$ is not very different from the case $q=3$ examined below. For $q=2$ we find, up to an integration constant, the following entropy:
\begin{equation}\label{S_q=2}
S_{q=2}^{\prime}=\frac{V\Omega_{3}\cosh\alpha}{4G}\left(r_{0}^{2}+2\gamma E_{p}^{-2}\right)\sqrt{r_{0}^{2}-\gamma E_{p}^{-2}}.
\end{equation}
We see that the entropy is positive and real for all $r_{0}>\sqrt{\gamma}/E_{p}$. Let us now examine a case where $q>n$. Let us choose the smallest of such integers; namely, $q=3$. Substituting in (\ref{modifiedS}), we find
\begin{equation}\label{S_q=3}
S_{q=3}^{\prime}=\frac{V\Omega_{3}\cosh\alpha}{4G}\left[\sqrt{r_{0}^{6}-\frac{\gamma r_{0}^{3}}{E_{p}^{3}}}+\frac{\gamma}{E_{p}^{3}}\ln{\left(\sqrt{r_{0}^{3}-\frac{\gamma}{E_{p}^{3}}}+r_{0}^{3/2}\right)}
-\frac{\gamma}{2E_{p}^{3}}\ln{\frac{\gamma}{E_{p}^{3}}}\right].
\end{equation}
We plot below the standard $5$-brane's entropy and rainbow $5$-brane's entropy side by side for comparison.
\begin{figure}[H]
\centering\includegraphics[scale=0.5]{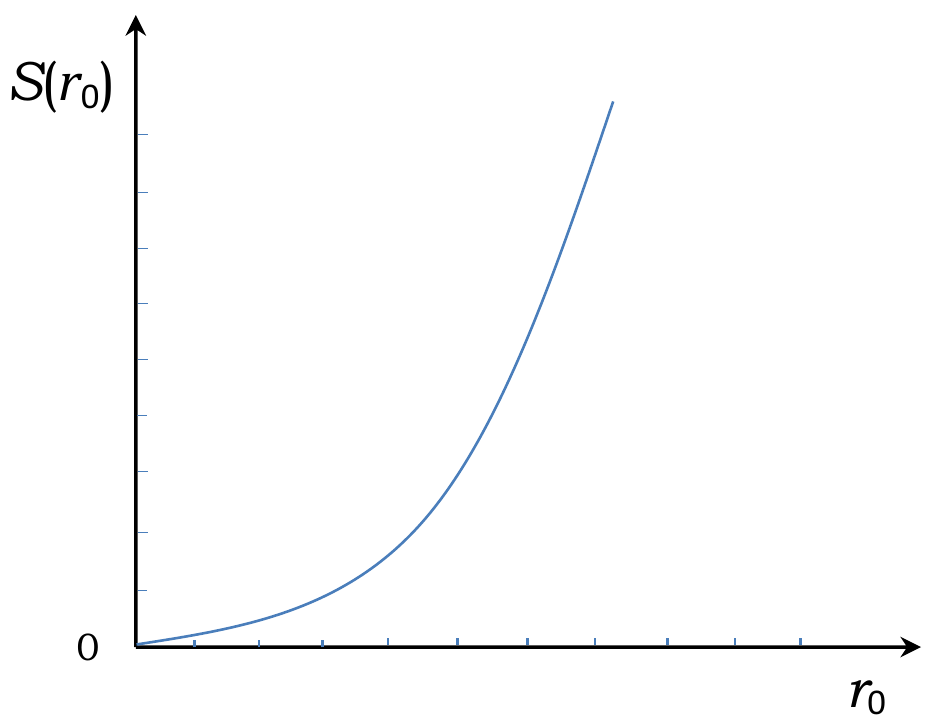}\qquad\includegraphics[scale=0.5]{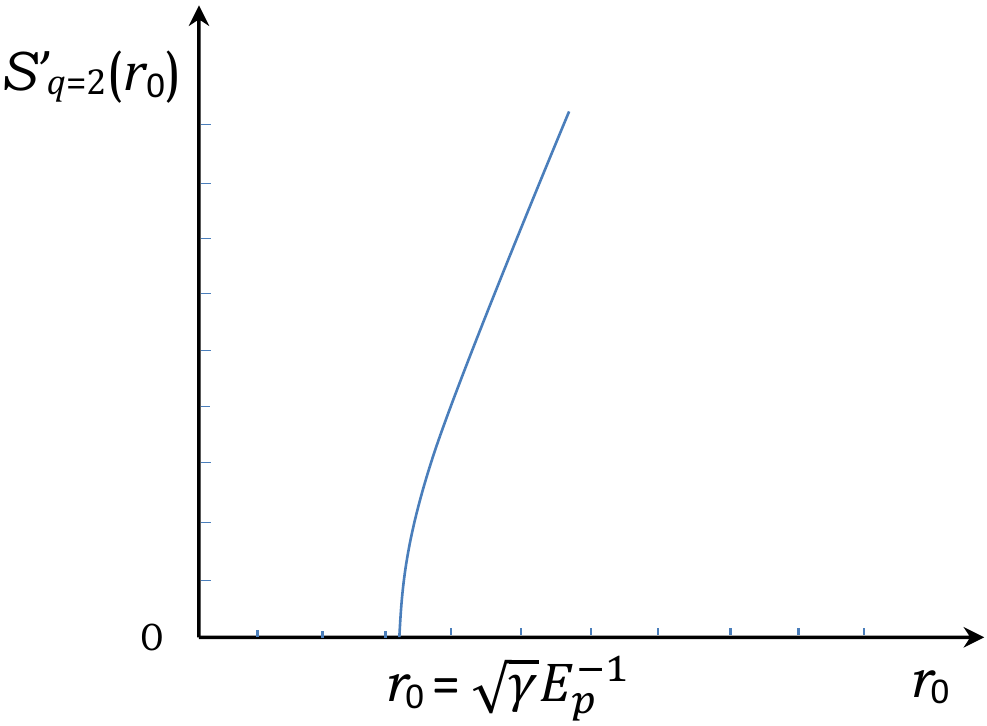}\qquad\includegraphics[scale=0.5]{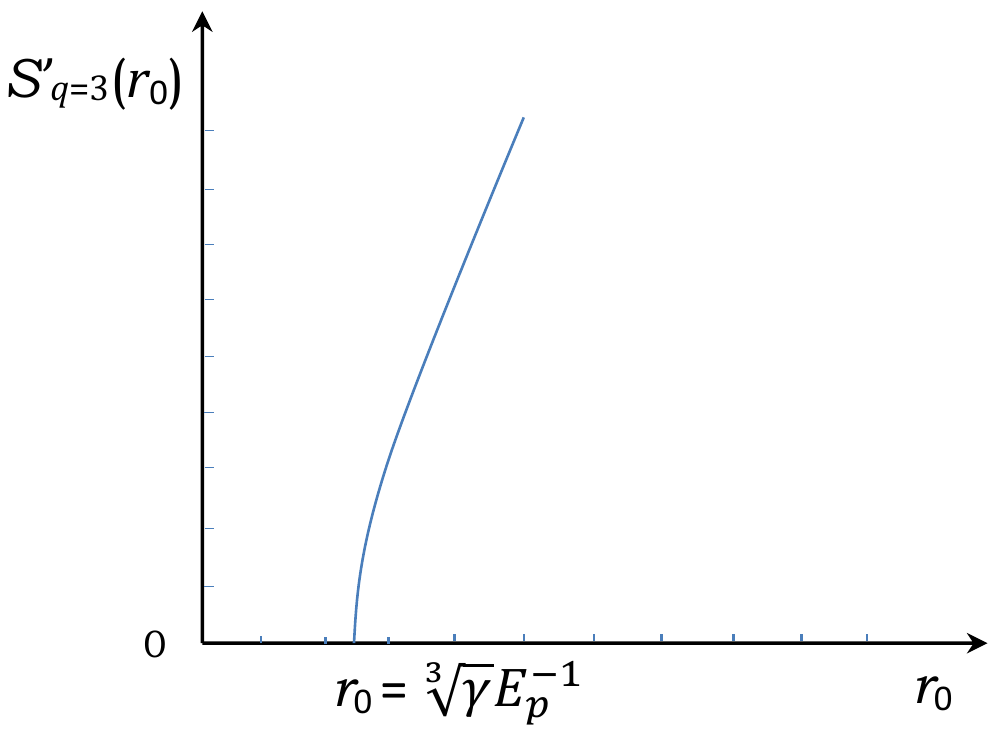}
\caption{Plot showing the entropies of a $5$-brane. The first graph represents the entropy of a standard $5$-brane vs. the radius $r_{0}$.
The second and third graphs represent the entropy of a rainbow $5$-brane, for the rainbow functions (\ref{RainFunc}) with $q=2$ and $q=3$, respectively.}.
\label{EntropyGraphs}
\end{figure}
We clearly see from Fig.~\ref{EntropyGraphs} that a major difference appears between the entropy of a standard $p$-brane in the first plot and the entropy of rainbow $p$-branes in the two other plots. While the entropy for the standard $p$-brane is defined for all positive values of the radius $r_{0}$ of the black brane, the entropies for rainbow branes with $q=2$ and $q=3$ cease to be defined for a specific value of the Schwarzschild radius $r_{0}$. For the case $q=2$, we see that values of the radius $r_{0}$ below $\sqrt{\gamma}E_{p}^{-1}$ are forbidden as they make the entropy become imaginary. For the case $q=3$, on the other hand, the forbidden values for the radius $r_{0}$ are found to be those values that are below $\sqrt[3]{\gamma}E_{p}^{-1}$. Notice that both these values could have been easily guessed simply by examining the square root in the rainbow functions (\ref{RainFunc}). We will come back to this important remark in Sec.~\ref{sec:5} below when we discuss the issue of being able to give a real physical meaning to remnants as they arise here.

Before we end this section, let us compute the heat capacity corresponding to the general entropy (\ref{modifiedS}). The result one finds is the following:
\begin{equation}\label{C'q=2}
C^{\prime}=T^{\prime} \frac{\partial S^{\prime}/{\partial r_0}}{\partial T^{\prime}/{\partial r_0}}=\frac{(n+1)Vr_{0}^{n+1}\Omega_{n+1}\cosh\alpha}{2G\left[(2+q)\gamma(E_{p}r_{0})^{-q}-2\right]}\sqrt{1-\gamma(E_{p}r_{0})^{-q}}.
\end{equation}
We note from this identity that the heat capacity of the black $p$-brane in Rainbow Gravity vanishes when $r_0=(\gamma)^{1/q}E_p^{-1}$.
This means that the black $p$-brane stops exchanging heat with the surrounding space even before reaching the minimal mass $M_{min}$ as given by (\ref{MinMass}). This is in agreement with the interpretation we made above of the constraint (\ref{MinMass}); namely, that the latter is predicting the existence of a remnant black brane. We plot below the variations of the heat capacity with the radius $r_{0}$. We plot here only the case $q=2$ because the same form is also what is recovered for all values of the integer $q$. As we can see from the plot the heat capacity exhibits a phase transition for the value $r_{0}=\sqrt{2\gamma}E_p^{-1}$  and ceases to be defined at that value for which the denominator in (\ref{C'q=2}) vanishes. The heat capacity also ceases to be defined when the square root in the numerator becomes imaginary for $r_{0}<\sqrt{\gamma}E_{0}^{-1}$, which constitutes thus the radius of the remnant black brane. This means that a black brane remnant will also exist due to this constraint from the heat capacity.
\begin{figure}[H]
\centering\includegraphics[scale=0.7]{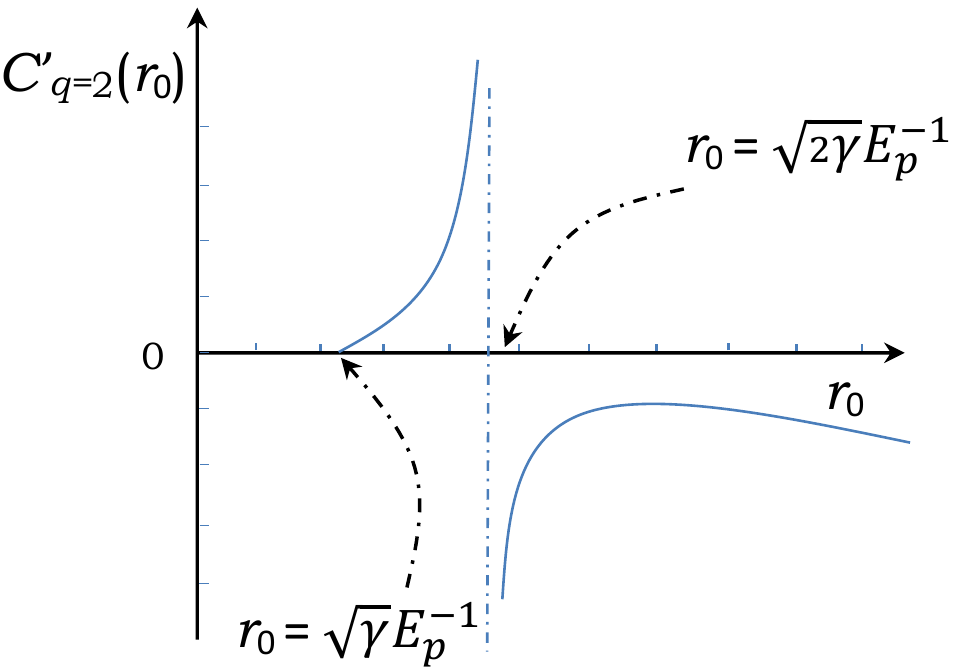}
\caption{The heat capacity of a rainbow $5$-brane corresponding to the rainbow functions (\ref{RainFunc}) with $q=2$.}
\label{HeatCapacity q=2}
\end{figure}

\section{Other Rainbow $p$-Branes}\label{sec:4}
It is instructive to study also the effect of other dispersion relations besides the form (\ref{MDR}) and verify
if the previous conclusions would still hold for these cases. Indeed, in the literature other forms of dispersion relations have
been introduced and studied. The rainbow functions for a MDR with constant velocity of light, can be written as  ~\cite{MagSmolin}
\begin{equation}\label{MDR2}
f(E/E_{p})=g(E/E_{p})=\frac{1}{1-\gamma E/E_{p}},
\end{equation}
where $\gamma$ is again a dimensionless factor of order unity and $E_{p}$ is the Planck energy.
A straightforward substitution in the formula $T'=f^{-1}(E/E_{p})g(E/E_{p})T$ reveals that, although the latter version of dispersion relation
modifies the BB metric, it does not introduce any new effect on the thermodynamics of the $p$-brane.

We would also like to point out that there are also experimental observations which suggest that the usual energy-momentum relation
might get modified in the UV limit.   The   Greisen-Zatsepin-Kuz'min
limit (GZK limit) as an upper limit on the energy of
cosmic rays can be used to study  quantum gravitational effects \cite{q5}.  It may be noted that  the Pierre
Auger Collaboration and the High Resolution Fly's Eye
(HiRes) experiment have reconfirmed earlier results of the
GZK cutoff \cite{q6}.  All these  observations suggest  the
the modification of the usual dispersion relation in the UV limit, and so
there is a strong experimental motivation for analyzing such a modified  dispersion relation.
The third and last version of MDR we would like to examine in this paper has been motivated by the hard spectra from gamma-ray
burster's~\cite{AmelinoCamelia:1997gz}:
\begin{equation}\label{MDR3}
f(E/E_{p})=\frac{e^{\gamma E/E_{p}}-1}{\gamma E/E_{p}}~~~\mathrm{and}~~~g(E/E_{p})=1.
\end{equation}

A substitution in the formula $T'=f^{-1}(E/E_{p})g(E/E_{p})T$ gives the following new temperature
\begin{align}
T^{\prime}&=\frac{n\gamma}{4\pi E_{p}r_{0}^{2}\cosh\alpha}\left(e^{\gamma(E_{p}r_{0})^{-1}}-1\right)^{-1},\label{mod2T}
\end{align}
where we have used again the lower bound of energy $E\geq r_{0}^{-1}$ to write the last result. Using the first law of thermodynamics $\mathrm{d}M=T'\mathrm{d}S'$ again with the mass given by (\ref{BraneMass}) and the above temperature (\ref{mod2T}), we find the following entropy:
\begin{equation}\label{modified2S}
S^{\prime}=\frac{(n+1)VE_{p}\Omega_{n+1}\cosh\alpha}{4\gamma G}\int \left(e^{\gamma(E_{p}r_{0})^{-1}}-1\right)r_0^{n+1}\mathrm{d}r_0.
\end{equation}
It is clear from this expression that for all $r_{0}$ we have a real and positive entropy.
This however does not imply that the Schwarzschild radius of the $p$-brane may take any value.
We can see this by computing the heat capacity as we done it above.
In fact, using again expression (\ref{BraneMass}) for the mass $M$ and the new temperature (\ref{mod2T}), we find the following result:

\begin{equation}\label{C'}
C^{\prime}=\frac{(n+1)VE_{p}^{2}r_{0}^{n+3}\Omega_{n+1}\cosh\alpha}{4\gamma G\left[2E_{p}r_{0}+(\gamma- 2E_{p}r_{0})e^{\gamma(E_{p}r_{0})^{-1}}\right]}
\left(e^{\gamma(E_{p}r_{0})^{-1}}-1\right)^{2}.
\end{equation}
\\
We plot below the variations of the heat capacity with the radius $r_{0}$.
\begin{figure}[H]
\centering\includegraphics[scale=0.7]{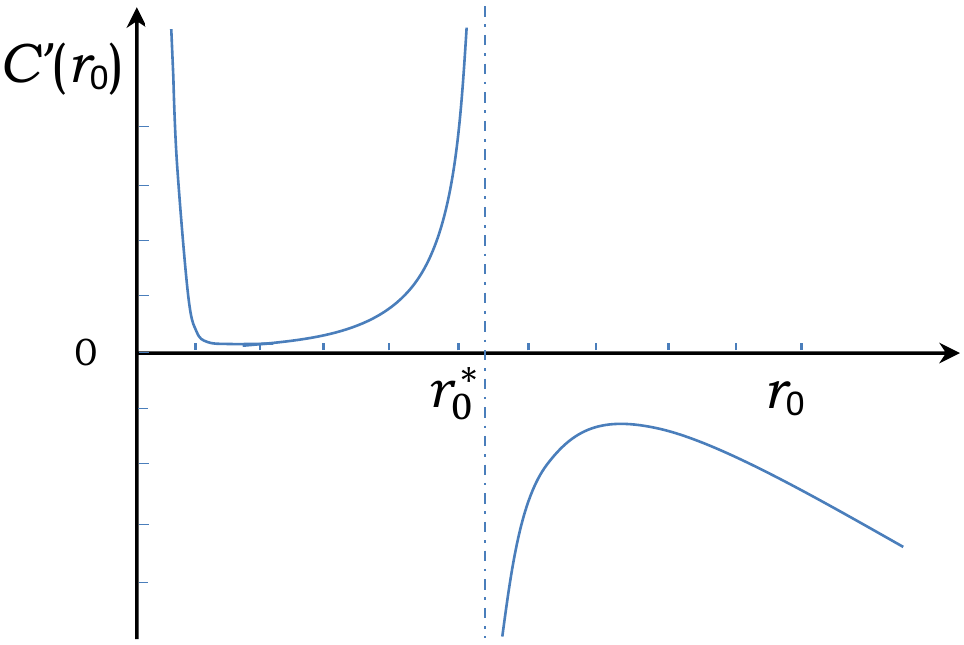}
\caption{The heat capacity of a rainbow $5$-brane corresponding to the rainbow functions (\ref{MDR3}).}
\label{HeatCapacity'}
\end{figure}
We see that the heat capacity goes to infinity and changes sign at $r_{0}=r_{0}^{*}$ for which the denominator in (\ref{C'}) vanishes.
This means that, in contrast to what we found for the rainbow functions (\ref{RainFunc}), the heat capacity does not vanish for any value of $r_{0}$ for a $p$-brane obeying a modified dispersion relation based on the rainbow functions (\ref{MDR3}).
This does not mean, however, that any value of $r_{0}$ is allowed, because the heat capacity, as we see from Fig.~\ref{HeatCapacity'},
changes sign at $r_{0}=r_{0}^{*}$ and ceases to be defined there.
This also implies that a phase transition occurs at this particular value of $r_{0}$ and, hence,
a remnant $p$-brane also exists under the version (\ref{MDR3}) of the MDR.
\section{On the reality of the remnants}\label{sec:5}
In this section we would like to discuss some subtleties concerning our conclusion leading to the existence of black remnants. In fact, the argument we used for the latter was heavily based on the fact that the entropy and heat capacity of the $p$-brane, as they come from formulas (\ref{S_q=2}), (\ref{S_q=3}) and (\ref{C'q=2}) in Sec.~\ref{sec:3} or formula (\ref{C'}) in Sec.~\ref{sec:4}, could not be defined below a certain value of the radius $r_{0}$ of the black brane. Therefore, one is naturally induced to wonder at what extent our physical conclusions could be trusted when the argument used to obtain them relies on a mathematical inconsistency that rises whenever one allows certain values for the radius $r_{0}$.

This issue becomes actually more serious when one notices that the inconsistency for the case of entropy rose from the rainbow functions (\ref{RainFunc}) which are themselves physically unacceptable whenever the ratio $E_{p}/\gamma^{{1/q}}$ is not bigger than the energy $E$ of the probe particle. Thus, a correlation between the inconsistencies in the rainbow functions and the impossibility of defining entropy and heat capacity for all values of $r_{0}$ is clearly apparent. Moreover, the absence of such an inconsistency in the rainbow functions (\ref{MDR2}) and (\ref{MDR3}) allowed us, as we saw, to have well-defined entropy expressions for all radiuses $r_{0}$. It is also certainly not a coincidence that the remnant mass, as it follows from (\ref{MinMass}), must be greater than $\gamma^{1/q}E_{p}$, i.e. that its Schwarzschild radius be greater than $\gamma^{1/q}E_{p}^{-1}$, while at the same time the energy $E$ of any probe particle must be smaller than $\gamma^{-1/q}E_{p}$ for the rainbow functions (\ref{RainFunc}) to be defined. This result, in fact, could have been guessed directly from the relation $E\geq r_{0}^{-1}$ we used without having to rely at all on the laws of thermodynamics. Thus, we would like to know if it really cannot be excluded that our results are simply due to mathematical artifacts.

It is actually very hard to come up with a solid physical argument that would remove completely any doubt about the validity of our conclusion. The best we could do is to evoke our result from section \ref{sec:4} where we were also led to conclude that there must be a remnant brane even though the entropy was defined for all values of $r_{0}$. We saw indeed that for the rainbow functions (\ref{MDR3}), which remain real for all values of the probe energy $E$, only the heat capacity, as given by (\ref{C'}), allowed us to generate constraints on the values of $r_{0}$. Therefore, if one category of rainbow functions yields a remnant brane without relying on any restriction carried by the rainbow functions themselves, we might very well assume that the constraints that led us to conclude in favor of the existence of remnants might be due, not only to the structure of the rainbow functions themselves, but also to the thermodynamics behind the rainbow branes.

In summary, then, we can say that while the existence of remnants could very well be simply due to mathematical artifacts, the need for consistency between our results in sections \ref{sec:3} and \ref{sec:4} could be used as further evidence in favor of the remnant interpretation we made of our results for entropy and heat capacity in this paper.

\section{Conclusion}

We have examined in this paper the effects of modified dispersion relations on the thermodynamics of $p$-branes.
We have adopted the same method for computing entropy using the mass of a black hole and its surface gravity.
By imposing a rainbow geometry on the black $p$-brane, we found that the entropy of the latter is modified accordingly.
In contrast to the standard $p$-brane thermodynamics, a rainbow black $p$-brane does not possess a well-defined entropy
below a specific value of its Schwarzschild radius. In other words, values of the black hole radius below a given minimum are not allowed for
a black $p$-brane. We interpreted this fact as implying the existence of a remnant brane. The computation of the heat capacity of such a brane
showed that the latter would indeed exhibit a phase transition if its radius decreases below the minimum allowed.

The previous conclusion does not, actually, apply for all rainbow $p$-branes. Indeed, by using two other different modified dispersion relations,
we found that for one MDR version a rainbow brane has exactly the same thermodynamics as a standard brane, whereas for another version of the MDR the black $p$-brane develops again a minimum radius and, hence, suggests the existence of a remnant. This was found, not from the expression of entropy which remains well-defined for all values of the radius, but from the computation of the heat capacity
which indeed changes sign, and even ceases to be defined, at a given value of the $p$-brane's Schwarzschild radius.
In fact, at that minimum radius, the heat capacity behaves just as if it would for a standard $p$-brane at zero Schwarzschild radius;
namely, it goes to $-\infty$ from the right. This fact allowed us, as we saw in Sec. \ref{sec:5}, to justify our interpretation of the inconsistency found for the first category of rainbow functions as the existence of a remnant $p$-brane and not only as a result of the constraints coming from the choice of the rainbow functions.

Next, we would like to note again the interesting fact that not all versions of the MDR lead to the same physics for rainbow branes. We have chosen to study in this paper only three of the main rainbow functions found in the literature; the first one being the most studied in the literature for having been implied by different approaches to quantum gravity. It is therefore interesting to go beyond these three families of functions and explore the more general case of rainbow functions $f(E/E_{p})$ and $g(E/E_{p})$ constrained only by the condition $\lim_{{E}\rightarrow0}f(E/E_{p})=1$ and $\lim_{{E}\rightarrow0}g(E/E_{p})=1$. However, the fact that the three families studied here yield different physical results might already constitute another important phenomenological criterion, beside the usual observational-based constraints, to favor one MDR version upon another. More rigorous investigations along these lines will be attempted in forthcoming works to gain a thorough understanding of rainbow branes and their role in the search for quantum gravity.

Finally, we would like to point out that in order to go from an energy-depend Hawking temperature (\ref{Rainsur}) to the radius-dependent temperature (\ref{modT}) we have used the standard Heisenberg uncertainty principle that allowed us to relate the lower bound of energy to the Schwarzschild radius. It is well-known, however, that having modified dispersion relations entails modified uncertainty relations,
i.e. the so-called generalized uncertainty principle (GUP). We therefore expect that the results obtained here will be modified when basing the
thermodynamics on the GUP. However, since the GUP brings tiny corrections to the usual Heisenberg uncertainty relations, we expect that also only tiny corrections will be brought to the different expressions found here for entropy and heat capacity of rainbow branes. The use of GUP would then certainly not alter the physical conclusions derived. A case study based on the full GUP will also be the subject of a forthcoming investigation.

\section*{Acknowledgments}
 We would like to thank Douglas  Smith for useful discussion. The research of Ahmed Farag Ali is supported by the STDF project 13858 and by Benha University (www.bu.edu.eg). The authors would like to thank the anonymous referee for constructive comments and suggestions that significantly helped to improve this paper.


\begin{thebibliography}{100}


\bibitem{AmelinoCamelia:1996pj}
  G.~Amelino-Camelia, J.~R.~Ellis, N.~E.~Mavromatos and D.~V.~Nanopoulos,
  Int.\ J.\ Mod.\ Phys.\ A {  12}, 607 (1997)

\bibitem{amerev}
  G.~Amelino-Camelia,
  Living Rev.\ Rel.\  {  16}, 5 (2013)


%
%
%

\bibitem{'tHooft:1996uc}
  G.~'t Hooft,
  Class.\ Quant.\ Grav.\  {  13}, 1023 (1996)

\bibitem{LIstring}
V. A. Kostelecky and S. Samuel, Phys. Rev. D {  39}, 683 (1989).

\bibitem{AmelinoCamelia:1997gz}
  G.~Amelino-Camelia, J.~R.~Ellis, N.~E.~Mavromatos, D.~V.~Nanopoulos and S.~Sarkar,
  Nature {  393}, 763 (1998)

\bibitem{Gambini:1998it}
  R.~Gambini and J.~Pullin,
  Phys.\ Rev.\ D {  59}, 124021 (1999)

\bibitem{Carroll:2001ws}
  S.~M.~Carroll, J.~A.~Harvey, V.~A.~Kostelecky, C.~D.~Lane and T.~Okamoto,
  Phys.\ Rev.\ Lett.\  {  87}, 141601 (2001)

\bibitem{AmelinoCamelia:1997jx}
  G.~Amelino-Camelia, J.~Lukierski and A.~Nowicki,
  Phys.\ Atom.\ Nucl.\  {  61}, 1811 (1998)
  [Yad.\ Fiz.\  {  61}, 1925 (1998)]

\bibitem{AmelinoCamelia:1999wk}
  G.~Amelino-Camelia, J.~Lukierski and A.~Nowicki,
  Int.\ J.\ Mod.\ Phys.\ A {  14}, 4575 (1999)

\bibitem{AmelinoCamelia:2002dx}
  G.~Amelino-Camelia,
  New J.\ Phys.\  {  6}, 188 (2004)

\bibitem{AmelinoCamelia:2000mn}
  G.~Amelino-Camelia,
  Int.\ J.\ Mod.\ Phys.\ D {  11}, 35 (2002)
  [gr-qc/0012051]; J.~Magueijo and L.~Smolin,
  Phys.\ Rev.\ D {  67}, 044017 (2003)

\bibitem{Magueijo:2002xx}
  J.~Magueijo and L.~Smolin,
  Class.\ Quant.\ Grav.\  {  21}, 1725 (2004)

\bibitem{Galan:2004st}
  P.~Galan and G.~A.~Mena Marugan,
  Phys.\ Rev.\ D {  70}, 124003 (2004);
  J.~Hackett,
  Class.\ Quant.\ Grav.\  {  23}, 3833 (2006); F.~Girelli, S.~Liberati and L.~Sindoni,
  Phys.\ Rev.\ D {  75}, 064015 (2007); C.~-Z.~Liu and J.~-Y.~Zhu,
  Gen.\ Rel.\ Grav.\  {  40}, 1899 (2008);
   H.~Li, Y.~Ling and X.~Han,
  Class.\ Quant.\ Grav.\  {  26}, 065004 (2009); R.~Garattini and G.~Mandanici,
  Phys.\ Rev.\ D {  85}, 023507 (2012); R. Garattini and F. S.N. Lobo, Phys.\ Rev.\ D {  85}, 024043 (2012);
  R. Garattini and G. Mandanici, Phys. Rev. D {  83}, 084021 (2011);
   J.~-J.~Peng and S.~-Q.~Wu,
  Gen.\ Rel.\ Grav.\  {  40}, 2619 (2008)


\bibitem{FRWRainbow}
 Y.~Ling,
  JCAP {  0708}, 017 (2007);
     Y.~Ling and Q.~Wu,
  Phys.\ Lett.\ B {  687}, 103 (2010)


\bibitem{Barrow:2013gia}
  J.~D.~Barrow and J.~Magueijo,
  arXiv:1310.2072 [astro-ph.CO];   G.~Amelino-Camelia, M.~Arzano, G.~Gubitosi and J.~Magueijo,
  Phys.\ Rev.\ D {  88}, 041303 (2013)


\bibitem{Awad:2013nxa}
  A.~Awad, A.~F.~Ali and B.~Majumder,
  JCAP {  1310}, 052 (2013)

\bibitem{HoravaPRD} P.~Horava, Phys. Rev. D 79,  084008 (2009).
\bibitem{HoravaPRL}  P.~Horava, Phys. Rev. Lett. 102, 161301 (2009).
\bibitem{A}  R. Gregory, S. L. Parameswaran, G. Tasinato and I. Zavala, JHEP1012, 047 (2010).
\bibitem{B}   P.~Burda, R.~Gregory and S.~Ross, JHEP  1411, 073 (2014).
\bibitem{ho} S. S. Gubser and A. Nellore,  Phys.Rev. D 80, 105007 (2009).
\bibitem{h1}  Y. C.  Ong and P.  Chen,  Phys. Rev. D 84, 104044 (2011).
\bibitem{h2}  M. Alishahiha and H.  Yavartanoo,   Class. Quant. Grav. 31, 095008 (2014).
\bibitem{oh} S. Kachru, N. Kundu, A. Saha, R. Samanta and S. P. Trivedi, JHEP1403, 074 (2014).
\bibitem{d} K.  Goldstein, N.  Iizuka, S. Kachru, S.  Prakash, S.  P. Trivedi and A.  Westphal,  JHEP 1010, 027 .
\bibitem{d1} G. Bertoldi, B. A. Burrington and A. W. Peet, Phys. Rev. D
82, 106013 (2010).
\bibitem{dh} M. Kord Zangeneh, A. Sheykhi and  M. H. Dehghani, Phys. Rev. D 92, 024050 (2015).
\bibitem{hd}  J. Tarrio and S. Vandoren, JHEP 1109, 017 (2011).

\bibitem{re}   R.~Garattini and E.~N.~Saridakis, Eur. Phys. J. C  75,   343 (2015).

\bibitem{FaizalJPA} M. Faizal,  J. Phys. A 44, 402001 (2011).
\bibitem{Carroll} S.~M. Carroll, J.~A. Harvey, V.~A. Kostelecky, C.~D. Lane and
T.~Okamoto, Phys. Rev. Lett.  87, 141601 (2001).
\bibitem{FaizalMPLA} M. Faizal,  Mod. Phys. Lett. A 27,  1250075  (2012).

\bibitem{st}N. Seiberg and E. Witten, JHEP 09, 032 (1999).
\bibitem{st1} Y.E. Cheung and M. Krogh, Nucl. Phys. B528, 185 (1998).
\bibitem{Amelino} G. Amelino-Camelia, Living Reviews in Relativity 5, 16
(2013).
\bibitem{Jacob} U. Jacob, F. Mercati, G. Amelino-Camelia and T. Piran, Phys. Rev. D 82, 084021
(2010).

\bibitem{58}  V. A. Kostelecky and S. Samuel,   Phys. Rev. D
39, 683 (1989).
\bibitem{59} V. A. Kostelecky and S. Samuel, Phys. Rev. D 40, 1886  (1989).

\bibitem{q5}K. Greisen, Phys. Rev. Lett. 16, 748 (1966)
\bibitem{q6} J.  Abraham  et  al.  (Pierre  Auger  Collaboration), Phys. Lett. B 685, 239 (2010)
\bibitem{q1} A. F. Ali and  M. Khalil,  Europhys. Lett.  110, 20009 (2015)
\bibitem{Lafrance:1994in}
  R.~Lafrance and R.~C.~Myers,
  Phys.\ Rev.\ D {  51}, 2584 (1995)

\bibitem{Mende:1992pm}
  P.~F.~Mende,
  hep-th/9210001.

\bibitem{Gross:1987ar}
  D.~J.~Gross and P.~F.~Mende,
  Nucl.\ Phys.\ B {  303}, 407 (1988)

\bibitem{Gubser:1998ex}
  S.~S.~Gubser,
  hep-th/9908004.

\bibitem{Polchinski:1996na}
  J.~Polchinski,
  hep-th/9611050

\bibitem{Horowitz:1996nw}
  G.~T.~Horowitz and J.~Polchinski,
  Phys.\ Rev.\ D {  55}, 6189 (1997)


\bibitem{Harmark} T. Harmark and N.A. Obers, JHEP01, 008 (2000)

\bibitem{Hawking:1974sw}
  S.~W.~Hawking,
  Commun.\ Math.\ Phys.\  {  43}, 199 (1975)
  [Erratum-ibid.\  {  46}, 206 (1976)];  S.~W.~Hawking,
  Phys.\ Rev.\ D {  13}, 191 (1976); J.~D.~Bekenstein,
  Lett.\ Nuovo Cim.\  {  4}, 737 (1972)
\bibitem{GRbooks}
Oyvind Gron, Einstein's general theory of relativity: with modern applications in cosmology. Springer, 2007;
R.M.Wald, General relativity, University of Chicago
Press; Chicago, (1984)

\bibitem{Bekenstein:1973ur}
  J.~D.~Bekenstein,
  Phys.\ Rev.\ D {  7}, 2333 (1973)

\bibitem{Adler}
 R.~J.~Adler, P.~Chen, D.~I.~Santiago,
  Gen.\ Rel.\ Grav.\  {  33}, 2101-2108 (2001)

\bibitem{Cavaglia:2003qk}
  M.~Cavaglia, S.~Das, R.~Maartens,
  Class.\ Quant.\ Grav.\  {  20}, L205-L212 (2003);
  M.~Cavaglia, S.~Das,
  Class.\ Quant.\ Grav.\  {  21}, 4511  (2004)


\bibitem{Niemeyer:2001xk}
  J.~C.~Niemeyer,
  Phys.\ Rev.\ D {  65}, 083505 (2002)
  [astro-ph/0111479]; A. Kempf, J.Phys. {  A 30}, 2093 (1997)



\bibitem{Eliasentropy0}
A.~J.~M.~Medved, E.~C.~Vagenas,
  Phys.\ Rev.\  D {  70}, 124021 (2004)

\bibitem{Eliasentropy1}
B.~Majumder,
  Phys.\ Lett.\ B {  703}, 402 (2011)

\bibitem{AmelinoCamelia:2004xx0}
  G.~Amelino-Camelia, M.~Arzano and A.~Procaccini,
  Phys.\ Rev.\ D {  70}, 107501 (2004);
E.M. Lifshitz, L.P. Pitaevskii and V.B. Berestetskii, Landau-Lifshitz Course of Theoretical
Physics, Volume 4: Quantum Electrodynamics, Reed Educational and Professional
Publishing, (1982)

\bibitem{AmelinoCamelia:2005ik}
  G.~Amelino-Camelia, M.~Arzano, Y.~Ling and G.~Mandanici,
  Class.\ Quant.\ Grav.\  {  23}, 2585 (2006)


\bibitem{Biswas:2011ar}
  T.~Biswas, E.~Gerwick, T.~Koivisto and A.~Mazumdar,
  Phys.\ Rev.\ Lett.\  {  108}, 031101 (2012)

\bibitem{Lu}
J.X.~Lu,
Phys.\ Rev. \ Lett.\ { 313}, 29 (1993)


\bibitem{Caio}
R-G.~Cai, Ru-K.~Su, P.~Yu,
 Phys. Lett. A 195, 307  (1994) n
\bibitem{Ali}
 A.~F.~Ali, M.~Faizal and M.~M.~Khalil,
  Nucl.\ Phys.\ B {  894}, 341 (2015)

\bibitem{MagSmolin}
 J. Magueijo and L. Smolin, Phys. Rev. Lett. 88, 190403 (2002)





\end{thebibliography}
\end{document}